\documentclass[aps,pre,twocolumn,groupedaddress]{revtex4}

\usepackage{graphicx}
\usepackage{bm}
\usepackage{amsmath}
\usepackage{subfigure}
\begin{document}

\title{Statistical Mechanics of Splay Flexoelectricity in Nematic
Liquid Crystals}

\author{Subas Dhakal}
\author{Jonathan V. Selinger}
\email{jselinge@kent.edu}
\affiliation{Liquid Crystal Institute, Kent State University, Kent, OH 44242}

\date{November 10, 2009}

\begin{abstract}
We develop a lattice model for the splay flexoelectric effect in nematic
liquid crystals.  In this model, each lattice site has a spin representing the
local molecular orientation, and the interaction between neighboring spins
represents pear-shaped molecules with shape polarity.  We perform Monte Carlo
simulations and mean-field calculations to find the behavior as a function of
interaction parameters, temperature, and applied electric field.  The
resulting phase diagram has three phases:  isotropic, nematic, and polar.  In
the nematic phase, there is a large splay flexoelectric effect, which diverges
as the system approaches the transition to the polar phase.  These results
show that flexoelectricity is a statistical phenomenon associated with the
onset of polar order.
\end{abstract}


\maketitle

\section{Introduction}

Flexoelectricity is a coupling between elastic deformation and electrostatic
polarization in a liquid crystalline medium.  In general, a splay or bend
deformation of the nematic director leads to an electrostatic polarization,
which can be observed as a macroscopic dipole moment of the liquid crystal.
Conversely, an applied electric field induces an electrostatic polarization,
which leads to a combination of splay and bend distortions in the nematic
director.  Since its discovery in 1969 by Meyer~\cite{meyer69}, the
flexoelectric effect has drawn great interest because of its possible
applications~\cite{Harden06,Harden08} in strain gauges, transducers,
actuators, micro power generator and electro-optical devices.

There have been many experimental and theoretical studies to determine the
flexoelectric coefficients of nematic liquid crystals~\cite{Straley76,%
Osipov84,Singh89,Somoza91,Zannoni91,Murthy93,Blinov98,Stelzer99,Billeter00,%
Blinov00,Berardi01,Zannoni01,Ferrarini01,Dewar05,Kapanowski07}, using a range
of different approaches.  For typical calamitic (rod-shaped) liquid crystals,
the splay and bend flexoelectric coefficients are in the range of 3--20 pC/m.
However, in recent experiments, Harden et al.~\cite{Harden06,Harden08} found
that bent-core liquid crystals have a surprisingly large bend flexoelectric
coefficient, up to 35 nC/m, roughly three orders of magnitude larger than the
typical value.  With this large bend flexoelectric coefficient, bent-core
liquid crystals may be practical materials for converting mechanical into
electrical energy.

For theoretical physics, a key question is how to explain the large
flexoelectric effect found in bent-core nematic liquid crystals, so that it
can be exploited for technological applications.  Our conjecture is that the
large flexoelectric effect is a statistical phenomenon associated with nearby
polar phase.  Near a polar phase, a nematic liquid crystal is on the verge of
developing spontaneous polar order, and hence any deformation of the director
should induce a large polar response.  To test this conjecture, we would like
to build a model with nematic and polar phases, and determine the behavior of
flexoelectric effect as a function of temperature above the nematic-polar
transition.  In this paper, we begin the study by investigating the splay
flexoelectric effect in a system of uniaxial pear-shaped molecules.  In a
subsequent paper, we will investigate the more complex case of bend
flexoelectricity in bent-core liquid crystals, as in the experiments.

\begin{figure*}
(a)\subfigure{\includegraphics[width=1.9in]{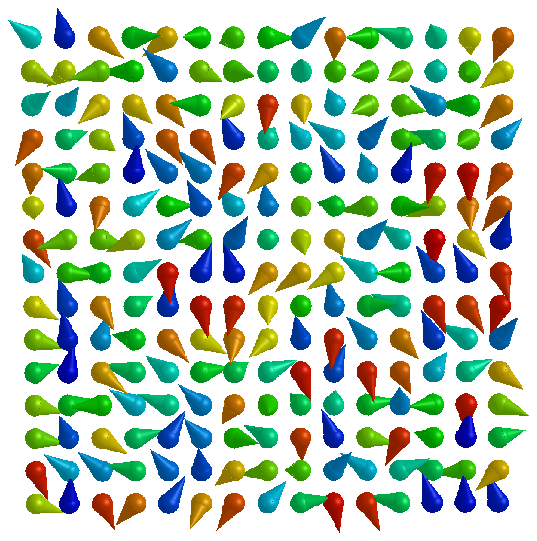}}
(b)\subfigure{\includegraphics[width=1.9in]{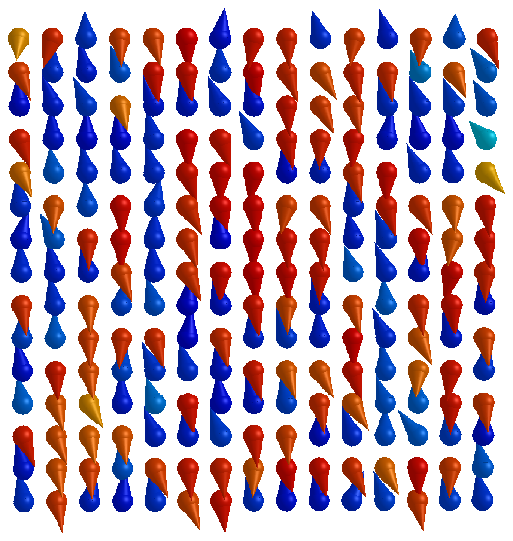}}
(c)\subfigure{\includegraphics[width=1.9in]{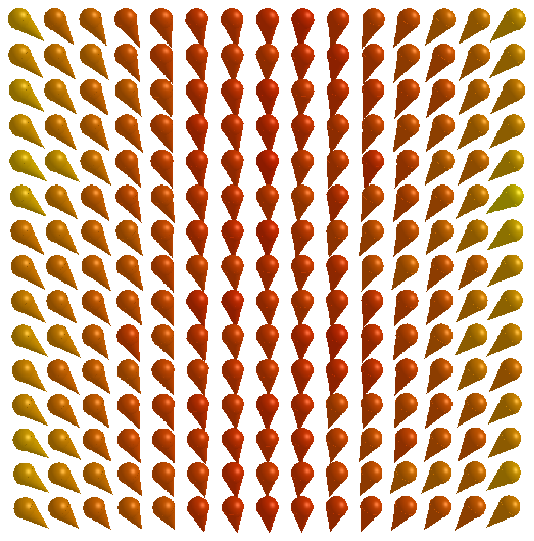}}
\caption{\label{snapshots}(Color online) Snapshots of the simulation results
in the three phases:  (a)~Isotropic.  (b)~Nematic.  (c)~Polar.  The software
V\_Sim is used~\cite{Vsim}, and the color of each molecule represents the
polar angle~$\theta$ away from the $z$-axis.}
\end{figure*}

To study the splay flexoelectric effect, we generalize the Lebwohl-Lasher
lattice model of nematic liquid crystals~\cite{Lebwohl72}.  In the original
Lebwohl-Lasher model, each lattice site $i$ has a spin $\mathbf{n}_i$, which
represents the local nematic director, with the symmetry
$\mathbf{n}_i\to-\mathbf{n}_i$.  In our generalization, the spins represent
the orientations of pear-shaped molecules, which do not have that symmetry.
For that reason, the interaction between neighboring spins includes three
terms---one term favoring nematic order, another term favoring polar order,
and a third term that couples polar order with splay of the nematic director.
With this interaction, we find a phase diagram with isotropic, nematic, and
polar phases, as illustrated in the snapshots of Fig.~\ref{snapshots}.  The
nematic phase has a flexoelectric effect, which increases as the system
approaches the polar phase.  Thus, this calculation demonstrates explicitly
that the flexoelectric effect is a collective, statistical phenomenon, which
is strongest near the transition to a phase with spontaneous polar order.

The plan of this paper is as follows.  In Sec.~II, we set up the theoretical
framework leading to the model interaction and discuss the relevant order
parameters.  In Sec. III, we present the Monte Carlo simulation methods and
results, for both the phase diagram and the flexoelectric effect.  In Sec.~IV
we present a mean-field theory for this model, and compare the simulation
results with the mean-field approximation.  Finally, in Sec.~V we discuss and
summarize the conclusions of this study.

\section{Model}

In this study, our goal is to simulate the splay flexoelectric effect in a
system of uniaxial pear-shaped molecules.  For these simulations, we construct
a lattice model that can represent both nematic and polar order.  In this
model, the local molecular orientation at lattice site $i$ is represented by a
unit vector $\hat{\bm{n}}_i$.  If the system has nematic order, then the
molecular orientations tend to be aligned along a preferred axis; i.e. there
is a nonzero order parameter
$\langle P_2(\hat{\bm{n}}_i\cdot\hat{\bm{d}})\rangle$, where $\hat{\bm{d}}$ is
the overall director and $P_2$ is the second Legendre polynomial.  If the
system has polar order, then the molecular orientations tend to point in a
particular direction; i.e. there is a nonzero order parameter
$\langle P_1(\hat{\bm{n}}_i\cdot\hat{\bm{d}})\rangle$, where $P_1$ is the
first Legendre polynomial.  Note that the system can have nematic order
without polar order, but it cannot have polar order without nematic order.

The lattice Hamiltonian must have four terms:  one term that favors nematic
order, one term that favors polar order, one term that gives a coupling
between polar order and an applied electric field, and a final term that gives
a coupling between polar order and splay of the nematic director.  The term
favoring nematic order can be written simply as
$-A(\hat{\bm{n}}_i\cdot\hat{\bm{n}}_j)^2$, summed over all pairs of
neighboring sites $i$ and $j$, as in the Lebwohl-Lasher
model~\cite{Lebwohl72}.  The term favoring polar order can be written even
more simply as $-B(\hat{\bm{n}}_i\cdot\hat{\bm{n}}_j)$, again summed over all
pairs of neighboring sites $i$ and $j$, as in the Heisenberg model of
magnetism.  The coupling between polar order and an applied electric field can
be written as $-\bm{E}\cdot\hat{\bm{n}}_i$, summed over $i$.

The coupling between polar order and nematic splay is somewhat more subtle.
For this coupling we need a lattice expression for the local splay between
neighboring sites $i$ and $j$.  Our expression for the local splay should
depend only on the nematic director, and hence it should be invariant under
the transformation $\hat{\bm{n}}\to-\hat{\bm{n}}$.  We cannot describe splay
by the scalar $\nabla\cdot\hat{\bm{n}}$, because it is not invariant under
that transformation.  Rather, we must describe splay by the vector
$\hat{\bm{n}}(\nabla\cdot\hat{\bm{n}})$, which has the correct symmetry.
In the following calculation, we let Latin letters refer to lattice sites and
Greek letters refer to directions.

On a continuum basis, the splay vector $\hat{\bm{n}}(\nabla\cdot\hat{\bm{n}})$
can be written in terms of the local nematic order tensor
$Q_{\alpha\beta}(\bm{r})$, or equivalently in terms of the dyad
$n_\alpha(\bm{r})n_\beta(\bm{r})$, as
\begin{eqnarray}
n_{\alpha}\partial_{\beta}n_{\beta}
&=&\frac{1}{2}\bigl[\partial_{\beta}(n_{\alpha}n_{\beta})
+(n_{\alpha}n_{\gamma})\partial_{\beta}(n_{\beta}n_{\gamma})\nonumber\\
&&\quad-(n_{\beta}n_{\gamma})\partial_{\beta}(n_{\alpha}n_{\gamma})\bigr].
\end{eqnarray}
Hence, a lattice approximation to the splay vector between sites $i$ and $j$
can be written as
\begin{eqnarray}
[n_{\alpha}\partial_{\beta}n_{\beta}]_{ij}
&=&\frac{1}{2}\biggl[r_{ij\beta}
(n_{j\alpha}n_{j\beta}-n_{i\alpha}n_{i\beta})\\
&&+\frac{n_{i\alpha}n_{i\gamma}+n_{j\alpha}n_{j\gamma}}{2}
r_{ij\beta}(n_{j\beta}n_{j\gamma}-n_{i\beta}n_{i\gamma})\nonumber\\
&&-\frac{n_{i\beta}n_{i\gamma}+n_{j\beta}n_{j\gamma}}{2}
r_{ij\beta}(n_{j\alpha}n_{j\gamma}-n_{i\alpha}n_{i\gamma})\biggr],
\nonumber
\end{eqnarray}
where $\hat{\bm{r}}_{ij}=\bm{r}_j-\bm{r}_i$ is the unit vector from site $i$
to $j$ on the lattice.  After some algebra, this expression simplifies to
\begin{eqnarray}
\label{splayvector}
[\hat{\bm{n}}(\nabla\cdot\hat{\bm{n}})]_{ij}&=&\frac{1}{2}\bigl[
 \hat{\bm{n}}_j(\hat{\bm{r}}_{ij}\cdot\hat{\bm{n}}_j)
-\hat{\bm{n}}_i(\hat{\bm{r}}_{ij}\cdot\hat{\bm{n}}_i)\nonumber\\
&&\quad+\hat{\bm{n}}_i(\hat{\bm{n}}_i\cdot\hat{\bm{n}}_j)
(\hat{\bm{r}}_{ij}\cdot\hat{\bm{n}}_j)\\
&&\quad-\hat{\bm{n}}_j(\hat{\bm{n}}_i\cdot\hat{\bm{n}}_j)
(\hat{\bm{r}}_{ij}\cdot\hat{\bm{n}}_i)\bigr].\nonumber
\end{eqnarray}
Note that this expression is invariant under the transformations
$\hat{\bm{n}}_i\to-\hat{\bm{n}}_i$, $\hat{\bm{n}}_j\to-\hat{\bm{n}}_j$, and
$i\leftrightarrow j$.

Now that we have found an expression for the local splay vector, we can couple
it with the local polar order.  The coupling term in the lattice Hamiltonian
can be written as the dot product of the splay between sites $i$ and $j$ with
the average polar order on these sites,
\begin{equation}
V_\mathrm{int}=-C[\hat{\bm{n}}(\nabla\cdot\hat{\bm{n}})]_{ij}\cdot
\frac{\hat{\bm{n}}_i+\hat{\bm{n}}_j}{2}.
\end{equation}
Simplifying with the use of Eq.~(\ref{splayvector}), the coupling term between
splay and polar order in the Hamiltonian is
\begin{equation}
V_\mathrm{int}=-C\left(\frac{1+\hat{\bm{n}}_i\cdot\hat{\bm{n}}_j}{2}\right)^2
\hat{\bm{r}}_{ij}\cdot\left(\hat{\bm{n}}_j-\hat{\bm{n}}_i\right).
\end{equation}
Combining all these terms, our final expression for the lattice Hamiltonian is
\begin{eqnarray}
\label{hamiltonian}
H&=&-\sum_{\langle i,j\rangle}\biggl[
A(\hat{\bm{n}}_i\cdot\hat{\bm{n}}_j)^2
+B(\hat{\bm{n}}_i\cdot\hat{\bm{n}}_j)\\
&&\quad+C\left(\frac{1+\hat{\bm{n}}_i\cdot\hat{\bm{n}}_j}{2}\right)^2
\hat{\bm{r}}_{ij}\cdot\left(\hat{\bm{n}}_j-\hat{\bm{n}}_i\right)\biggr]
-\sum_i \bm{E}\cdot\hat{\bm{n}}_i .\nonumber
\end{eqnarray}

At this point, we want to use this lattice Hamiltonian to calculate the
nematic order parameter $\langle P_2\rangle$, the polar order parameter
$\langle P_1\rangle$, and the average splay vector
${\langle\hat{\bm{n}}(\nabla\cdot\hat{\bm{n}})\rangle}$ as functions of the
parameters $A$, $B$, and $C$ and the electric field $\bm{E}$.  In the
following sections, we will do this calculation through Monte Carlo
simulations and mean-field theory.

\section{Monte Carlo simulation}

As a first step in exploring this model, we carry out Monte Carlo simulations
of a system of pear-like molecules interacting with the lattice Hamiltonian of
Eq.~(\ref{hamiltonian}). In these simulations, we use a simple cubic lattice
of size
$20\times20\times20$.  When an electric field is applied, it is in the $z$
direction, so that the molecules tend to align along $z$, with splay in the
$x$ and $y$ directions.  The lattice has periodic boundary conditions in $z$,
but free boundaries in $x$ and $y$, so that it can form splay in those
directions.

The usual Metropolis algorithm was used for lattice updates.  In each Monte
Carlo step, a lattice site is chosen randomly, its orientation is changed
slightly, and the change in energy $\Delta E$ is calculated.  If
$\Delta E < 0$ the move is accepted, and if $\Delta E > 0$ the move is
accepted with probability $\exp(-\Delta E/k_B T)$.  Starting from the
high-temperature isotropic phase, the system is cooled down slowly with
temperature steps of $\Delta T=0.02$.  The final configuration at each
temperature is taken as the initial configuration for the next lower
temperature. Typical runs take about $10^5$ steps to come to equilibrium,
while runs near phase transitions take about $6\times 10^5$ steps.  The
nematic and polar order parameters and the splay vector are calculated and
time-averaged during the production cycle.

The nematic order parameter $\langle P_2 \rangle$ is calculated by the usual
method using the $3D$ nematic order tensor
\begin{equation}
Q_{\alpha\beta}=\frac{1}{N}\sum_{i=1}^N\left(\frac{3}{2}n_{i\alpha}n_{i\beta}
-\frac{1}{2}\delta_{\alpha\beta}\right).
\end{equation}
where $\alpha$ and $\beta=x, y, z$, and  $N$ is the total number of lattice
sites.  The largest eigenvalue of this order tensor corresponds to
$\langle P_2 \rangle$.

To calculate the polar order parameter $\langle P_1 \rangle$, we assume that
polar order is oriented along the same axis as nematic order, as expected for
uniaxial molecules.  The eigenvector corresponding to the largest eigenvalue
of the nematic order tensor $Q_{\alpha\beta}$ is the instantaneous director
$\hat{\bm{d}}$.  Hence, the polar order parameter is calculated as the average
dot product of the director with the molecular orientation,
\begin{equation}
\langle P_1 \rangle=\frac{1}{N}\sum_{i=1}^N \hat{\bm{d}}\cdot\hat{\bm{n}}_i.
\end{equation}

The splay vector is calculated from Eq.~(\ref{splayvector}), averaged over
the four bonds
in the $(x,y)$ plane.  The magnitude of this vector gives
the average angle between the molecular orientations on neighboring lattice
sites.  For that reason, we report this magnitude as
$\langle\Delta\theta\rangle$.

\begin{figure*}
(a)\subfigure{\includegraphics[width=3.0in]{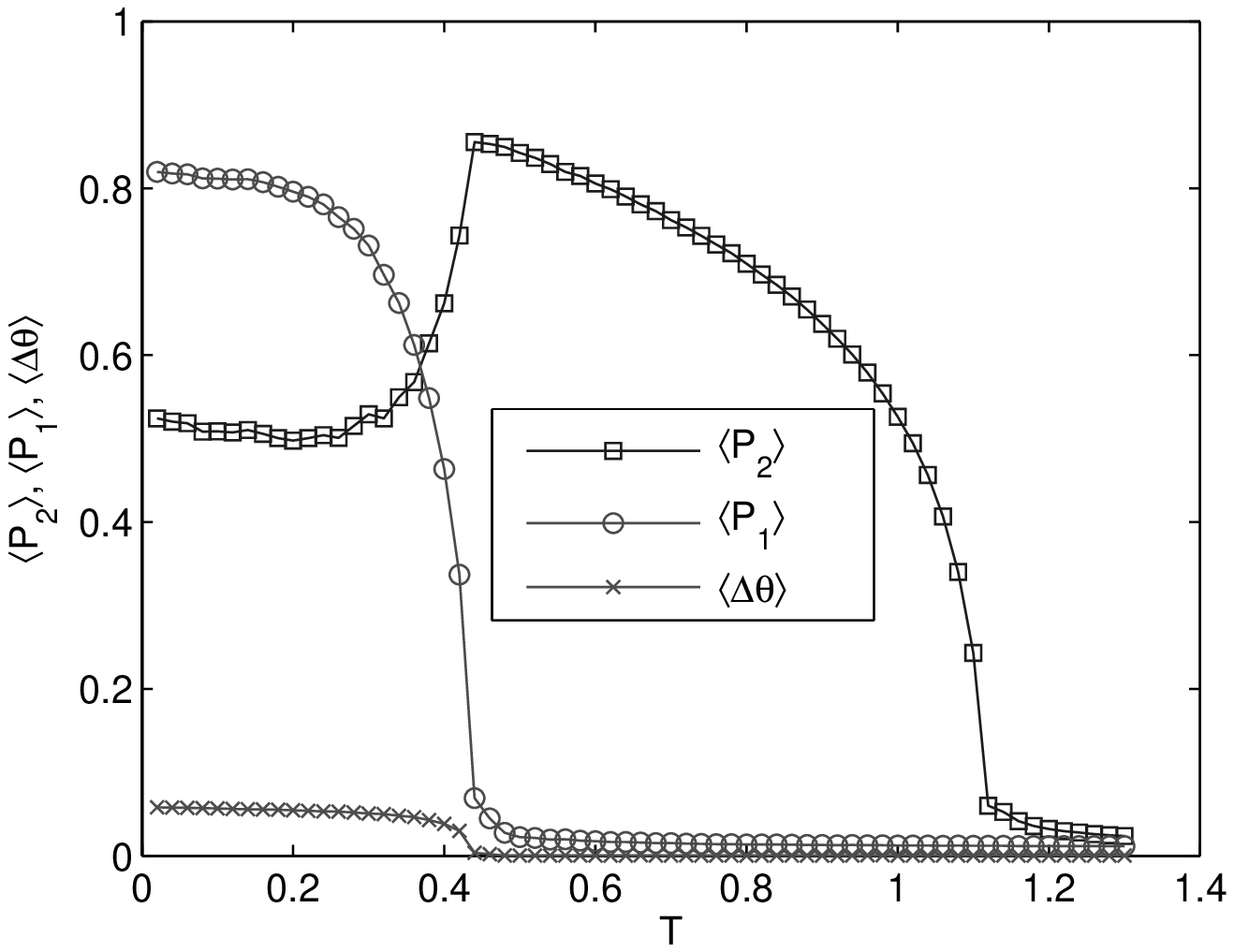}}
(b)\subfigure{\includegraphics[width=3.0in]{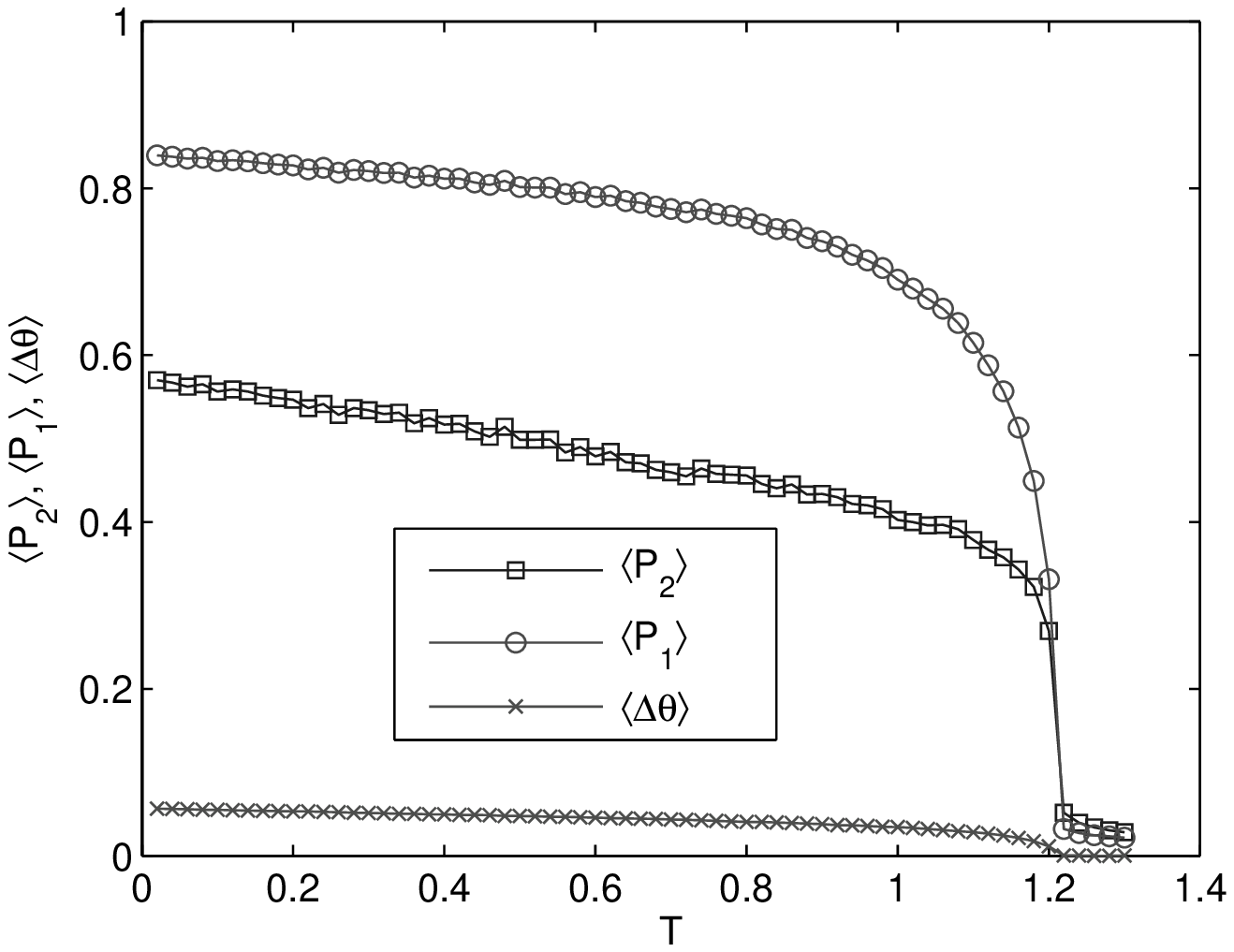}}
(c)\subfigure{\includegraphics[width=3.0in]{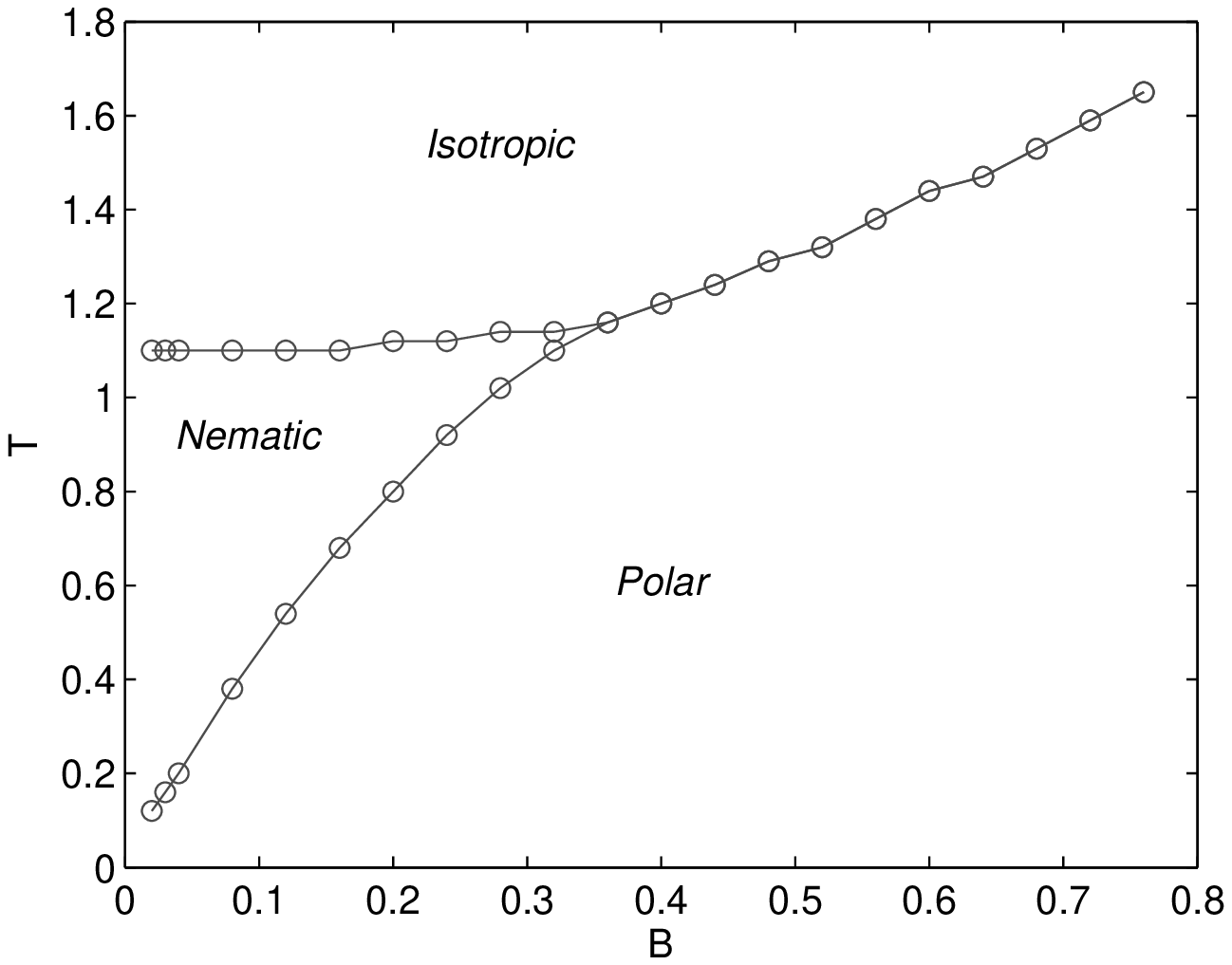}}
(d)\subfigure{\includegraphics[width=3.0in]{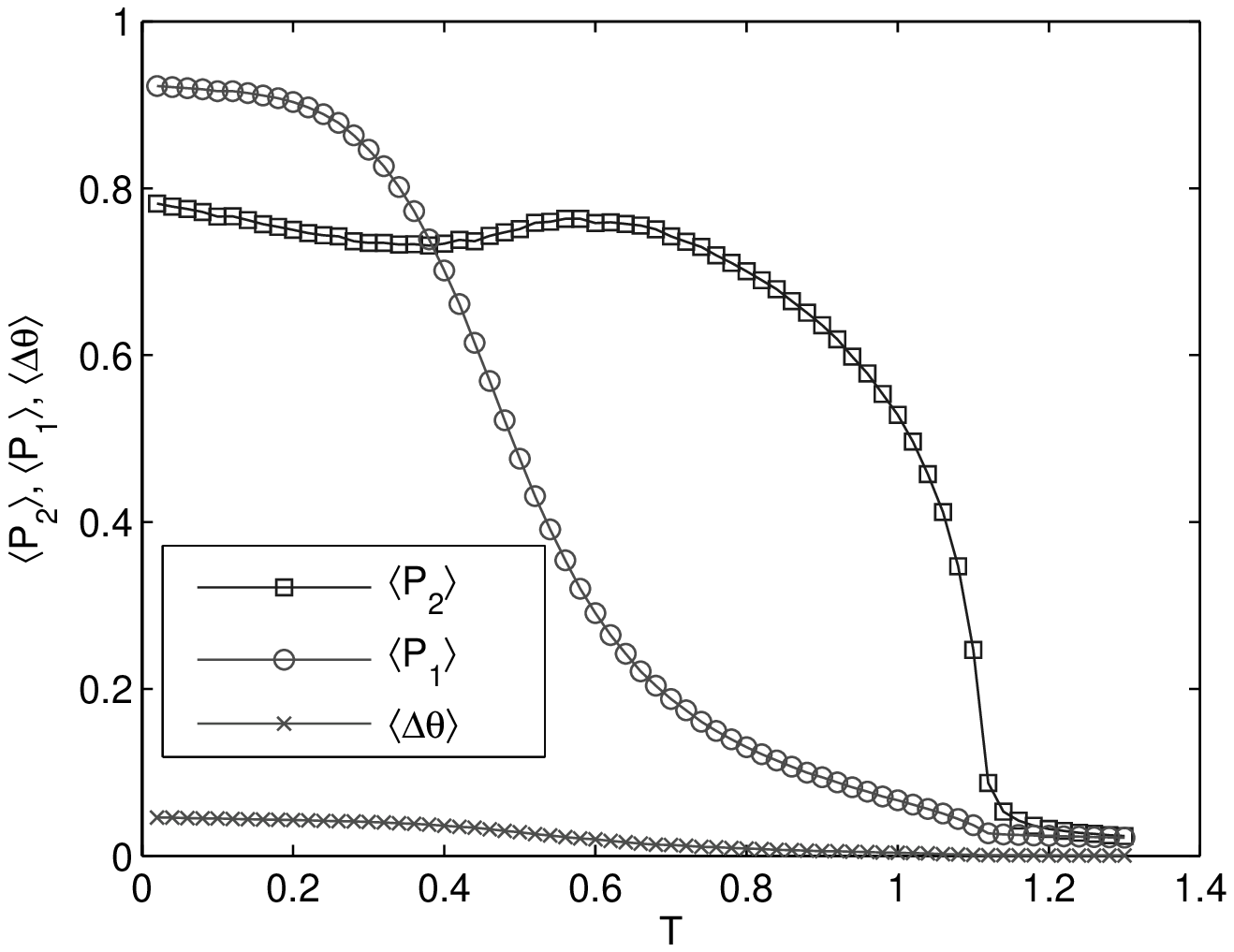}}
\caption{\label{orderparameterplots}Monte Carlo simulation results for the
order parameters $\langle P_1 \rangle$, $\langle P_2 \rangle$, and
$\langle\Delta\theta\rangle$ as functions of temperature $T$, for different
values of the interaction parameters chosen to show various types of
transitions:  (a)~Zero-field results for $A=1.5$, $B=0.09$, $C=0.3$, showing
the isotropic-nematic and nematic-polar transitions.  (b)~Zero-field results
for $A=1.5$, $B=0.4$,  $C=0.3$, showing the direct isotropic-polar transition.
(c)~Phase diagram for zero field.  (d)~Simulation with applied electric field
$E=0.06$, for the same parameters as in part~(a), showing the induced polar
order and splay in the nematic phase.}
\end{figure*}

Figure~\ref{orderparameterplots} shows plots of the order parameters
$\langle P_2 \rangle$, $\langle P_1 \rangle$, and $\langle\Delta\theta\rangle$
as functions of temperature for several values of the interaction parameters.
In Fig.~\ref{orderparameterplots}(a), for a small polar coupling $B$ and no
applied electric field, we see an isotropic-nematic transition at high
temperature followed by a nematic-polar transition at low temperature.  At the
isotropic-nematic transition, the nematic order parameter goes from zero to a
nonzero value. Here the transition is rounded by finite-size effects; we would
expect a sharp first-order transition for an infinite system.  Throughout the
nematic temperature range, the polar order parameter and splay are both zero.
At the nematic-polar transition, the polar order parameter becomes nonzero,
and this polar order induces an accompanying splay.  The nematic order
parameter decreases as the system moves into the polar phase, because the
splayed molecular orientation partially averages out the alignment, as shown
in the snapshot of Fig.~\ref{snapshots}(c).

In Fig.~\ref{orderparameterplots}(b), for a larger polar coupling $B$, we see
a direct transition from the isotropic to the polar phase, with no intervening
nematic phase.  In this case, the nematic and polar order parameters both
become nonzero at the same transition temperature.  Once again, the polar
order induces a splay, which inhibits the growth of the nematic order
parameter.

The simulation results for the phase diagram are shown in
Fig.~\ref{orderparameterplots}(c).  In this phase diagram, the vertical axis
shows temperature while the horizontal axis shows the polar coupling $B$ for a
constant nematic coupling $A$.  For small $B$ the phase diagram shows
isotropic,
nematic, and polar phases, with a nematic range that decreases as $B$
increases.  At a sufficiently large value of $B$, the nematic phase
disappears and there is a direct transition from isotropic to polar.  Note
that this phase diagram is quite similar to the phase diagram found in recent
work on the 2D isotropic, tetratic, and nematic phases~\cite{Geng09}.

The question is now:  What happens to the nematic phase when an electric field
is applied?  To answer this question, Fig.~\ref{orderparameterplots}(d) shows
the simulation results for the same parameters as
Fig.~\ref{orderparameterplots}(a), but in the presence of a small electric
field.  In the high-temperature isotropic phase, the field induces some polar
and nematic order, but this effect is very small.  However, in the nematic
phase, the field induces a more substantial polar order, and that polar order
induces a splay in the nematic director, i.e. a converse flexoelectric effect.
Both the polar order and the splay are quite temperature-dependent, increasing
as the system approaches the nematic-polar transition temperature, as would be
expected for a divergent susceptibility above a second-order transition.  The
nematic-polar transition is rounded off by the applied field, and the polar
order parameter and splay saturate in the low-temperature polar phase.

\begin{figure}
\includegraphics[width=3.0in]{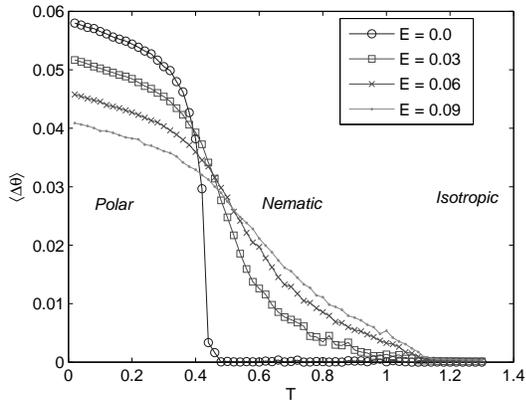}
\caption{Variation of splay as a function of temperature for several values of
the applied electric field.}
\label{Edependence}
\end{figure}

To provide further insight into the effect of an applied electric field,
Fig.~\ref{Edependence} shows the splay as a function of temperature for
several values of the field.  Within the nematic phase, the splay increases as
the electric field increases, as expected for the converse flexoelectric
effect. This trend is reasonable because an increasing electric field enhances
polar order.  For small field the splay is quite sensitive to temperature, but
for large field it becomes less temperature-dependent, as the induced polar
order grows larger and approaches saturation.  In the low-temperature polar
phase, the splay shows the opposite trend with electric field; it now
decreases as the field increases.  This trend is reasonable because an
increasing electric field cannot enhance the polar order, which is already
saturated; it only aligns the direction of polar order.  This alignment
reduces the induced splay, since splay necessarily involves some misalignment
of the molecular orientation.

\begin{figure}
\includegraphics[width=3.0in]{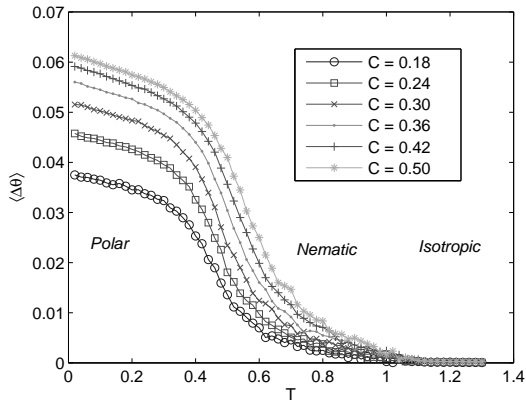}
\caption{Variation of splay as a function of temperature for several values of
the interaction coefficient $C$, with a small field $E=0.06$.}
\label{Cdependence}
\end{figure}

For comparison, Fig.~\ref{Cdependence} presents a plot of splay as a
function of temperature for several values of the interaction coefficient $C$.
This graph shows that the splay increases as $C$ increases, over the full
temperature range, in all phases.  This result is reasonable because the
coefficient $C$ represents the flexoelectric coupling between polar order and
splay.

As a final point, note that the behavior presented here can only occur in the
limit of small splay ${\Delta\theta\alt\pi/N}$, where $N$ is the system size.
In the opposite limit ${\Delta\theta\agt\pi/N}$, the system is too large for
one single splay from side to side.  Instead, it must break up into modulated
structures consisting of regions of splay separated by domain walls.  These
modulated structures might be splay stripes or even more complex two- or
three-dimensional arrangements of splay cells~\cite{Kamien01}.  We have
observed such modulated structures in our simulations, but we have not
explored them in detail because they are not likely to occur in experiments,
where the magnitude of splay is generally small.

\section{Mean-field calculation}

In this section we discuss two approximate analytic approaches to solve
the problem.  First, we map the interaction onto an Ising model, and use this
Ising model to calculate the splay and polar order as functions of temperature
and electric field.  Second, we present a more general mean-field calculation
with full rotational degrees of freedom, and use it to calculate the full
phase diagram with isotropic, nematic, and polar phases.

\subsection{Ising model}

For a simple Ising-type model of the splay flexoelectric effect, we suppose
that the system has well-defined nematic order, with variable amounts of splay
and polar order.  Consider a particular site $i$ surrounded by six nearest
neighbors on a cubic lattice.  We suppose that site $i$ has its director along
the $z$-axis, as do the two neighbors above and below, while the four nearest
neighbors in the $xy$-plane have directors that are splayed outward by a small
angle $\Delta\theta$.  The polar order at any site $j$ is represented by an
Ising spin variable $\sigma_j=\pm1$, which indicates whether the molecular
orientation is pointing up or down along the local director.  Thus, the
central site $i$ has the molecular orientation
$\hat{\bm{n}}_i=\sigma_i(0,0,1)$, while the six neighbors have the
orientations
$\hat{\bm{n}}_\mathrm{+x}=\sigma_\mathrm{+x}
(\sin\Delta\theta,0,\cos\Delta\theta)$,
$\hat{\bm{n}}_\mathrm{-x}=\sigma_\mathrm{-x}
(-\sin\Delta\theta,0,\cos\Delta\theta)$,
$\hat{\bm{n}}_\mathrm{+y}=\sigma_\mathrm{+y}
(0,\sin\Delta\theta,\cos\Delta\theta)$,
$\hat{\bm{n}}_\mathrm{-y}=\sigma_\mathrm{-y}
(0,-\sin\Delta\theta,\cos\Delta\theta)$,
$\hat{\bm{n}}_\mathrm{+z}=\sigma_\mathrm{+z}(0,0,1)$, and
$\hat{\bm{n}}_\mathrm{-z}=\sigma_\mathrm{-z}(0,0,1)$.

We now substitute these expressions for the molecular orientations into the
lattice Hamiltonian of Eq.~(\ref{hamiltonian}).  As usual in mean-field
theory, we assume that all the neighbors of site $i$ have polar order given by
$\langle\sigma_j\rangle=M$.  (This quantity is called $P_1$ in the other
sections; here we use the symbol $M$ to emphasize the analogy with the Ising
magnetization.)  The mean potential experienced by site $i$, expanded to
second order in the small splay $\Delta\theta$, is then
\begin{eqnarray}
V_\mathrm{mean}&=&-A(6-4(\Delta\theta)^2)
-B M \sigma_i(6-2(\Delta\theta)^2)\nonumber\\
&&-2C\Delta\theta(\sigma_i+M)-E\sigma_i .
\end{eqnarray}
Hence, the effective field acting on the Ising spin $\sigma_i$ is
\begin{equation}
E_\mathrm{eff}=E + 6 B M + 2 C \Delta\theta - 2 B M (\Delta\theta)^2 .
\end{equation}
As a result, the polar order parameter must satisfy the self-consistency
equation
\begin{eqnarray}
\label{selfconsistency}
M&=&\langle\sigma_i\rangle=\tanh\left(\frac{E_\mathrm{eff}}{k_B T}\right)\\
&=& \tanh\left(\frac{E + 6 B M + 2 C \Delta\theta - 2 B M (\Delta\theta)^2}%
{k_B T}\right).\nonumber
\end{eqnarray}
Furthermore, minimization of the mean potential over the splay $\Delta\theta$
gives
\begin{equation}
\label{deltathetamin}
\Delta\theta=\frac{C M}{2 A + B M^2}.
\end{equation}
Solving Eqs.~(\ref{selfconsistency}) and~(\ref{deltathetamin}) simultaneously
gives the equilibrium values of the splay $\Delta\theta$ and polar order $M$,
as functions of electric field $E$, temperature $T$, and energetic parameters
$A$, $B$, and $C$.

To calculate the response to an electric field in the nematic phase, we assume
that $E$, $M$, and $\Delta\theta$ are all small, and expand
Eqs.~(\ref{selfconsistency}) and~(\ref{deltathetamin}) to linear order in this
quantities.  These equations imply
\begin{eqnarray}
\label{polarresponse}
M&=&\frac{E}{k_B T - (6B+C^2/A)},\\
\label{converseflexoresponse}
\Delta\theta&=&\frac{C E}{2A[k_B T - (6B+C^2/A)]} .
\end{eqnarray}
Note that Eq.~(\ref{polarresponse}) gives the polar order parameter induced by
an applied electric field, while Eq.~(\ref{converseflexoresponse}) gives the
converse flexoelectric effect induced by the field.  Both of these responses
increase as the temperature decreases toward the second-order nematic-polar
transition at the temperature
\begin{equation}
k_B T_{NP} = 6B+\frac{C^2}{A}.
\end{equation}
At this transition, they diverge as $(T-T_{NP})^{-\gamma}$, with critical
exponent $\gamma=1$, as expected for the susceptibility to an applied field,
in mean-field theory for the Ising model.

\begin{figure*}
(a)\subfigure{\includegraphics[width=3.0in]{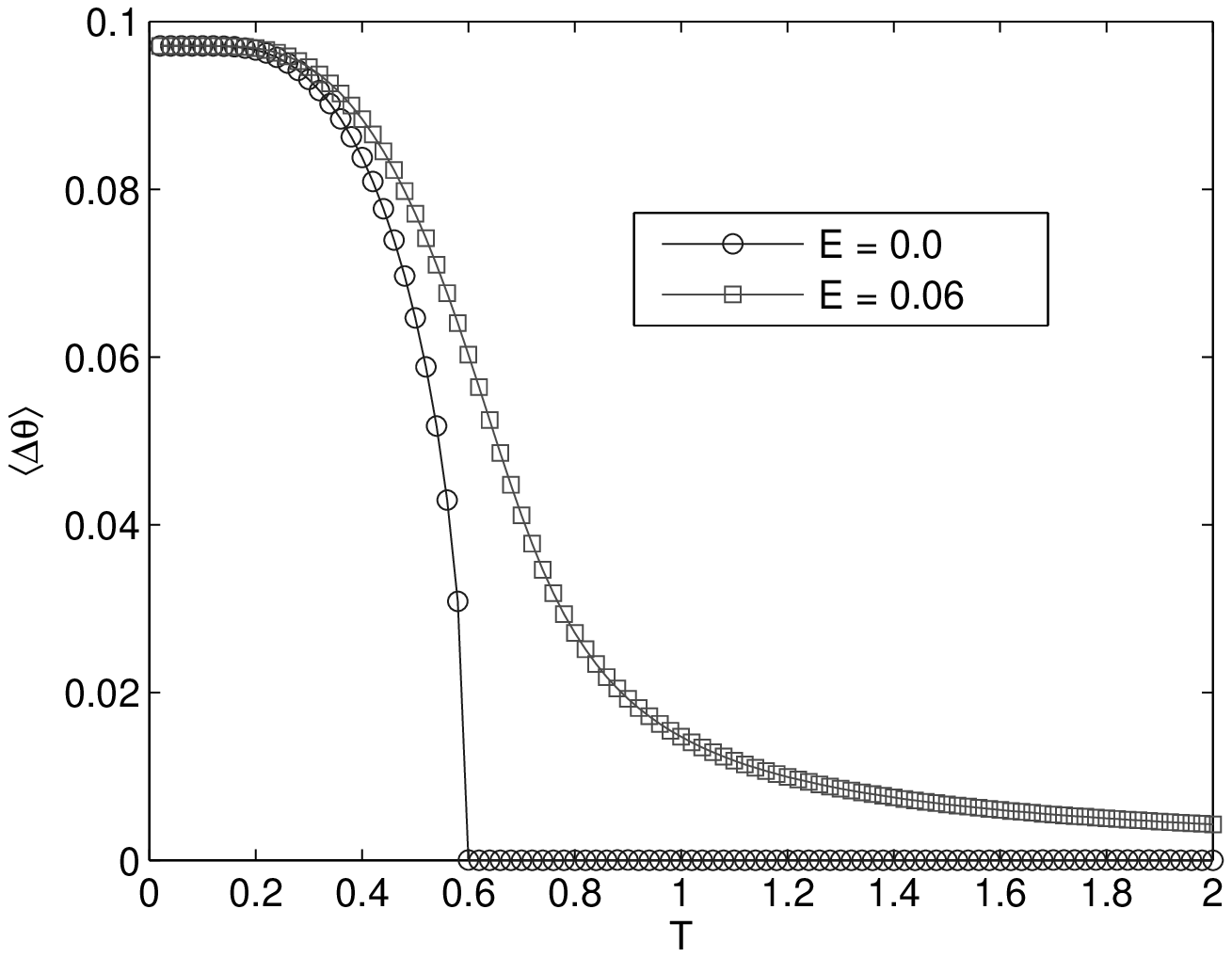}}
(b)\subfigure{\includegraphics[width=3.0in]{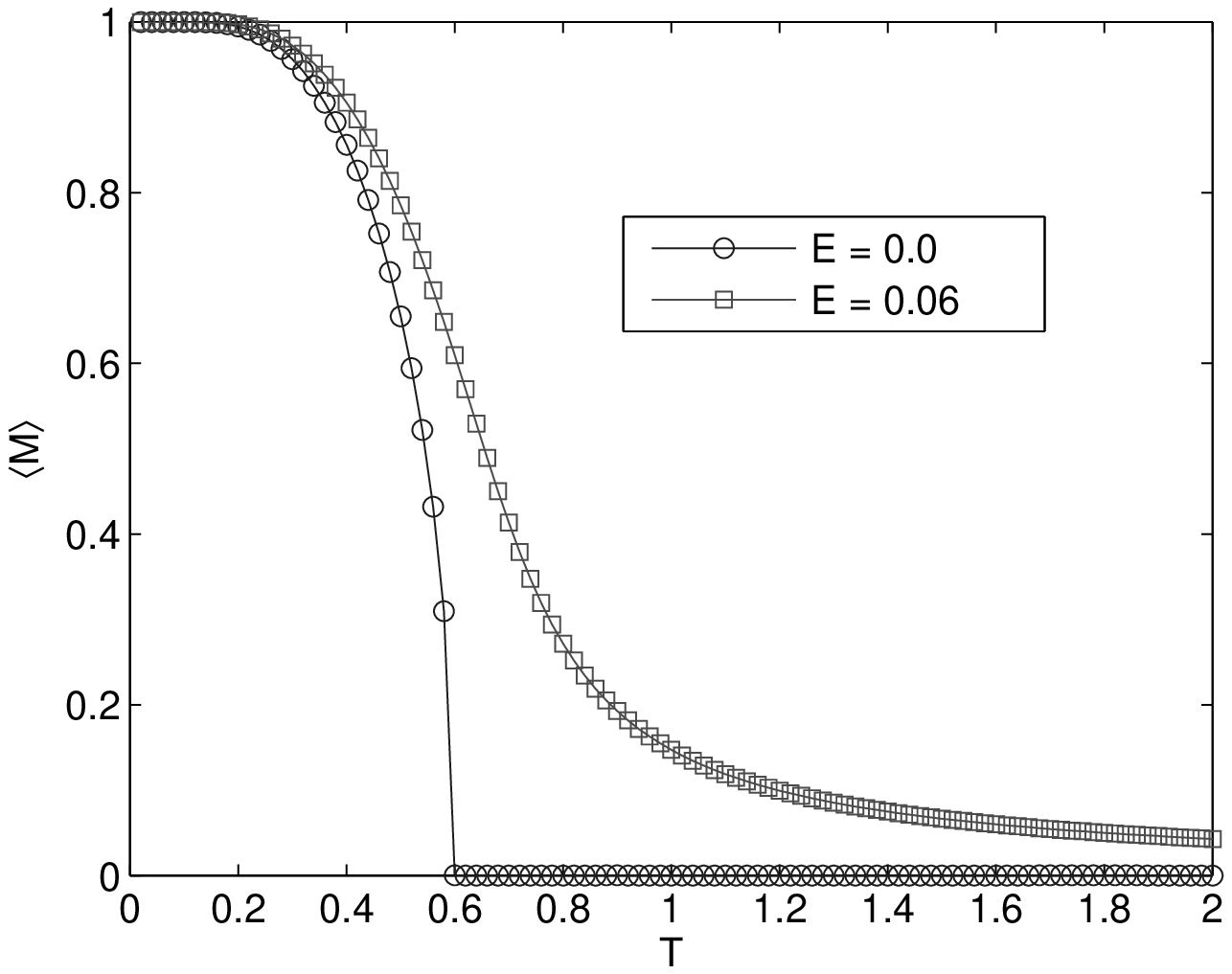}}
\caption{\label{numericalising} Numerical mean-field calculations for the
Ising mapping, showing the splay and polar order as functions of
temperature~$T$, for parameters $A=1.5$, $B=0.09$, and $C=0.3$, for zero and
nonzero electric field.}
\end{figure*}

For a more precise calculation, we solve Eqs.~(\ref{selfconsistency})
and~(\ref{deltathetamin}) numerically as functions of temperature and field.
The numerical results for splay $\Delta\theta$ and polar order $M$ are shown
in Fig.~\ref{numericalising}.  As in the approximate analytic calculation
above, we see a second-order nematic-polar transition.  The high-temperature
nematic phase has no polar order or splay without a field, but an applied
field induces both of these quantities.  By contrast, the low-temperature
polar phase has both spontaneous polar order and spontaneous splay, and they
both increase moderately when a field is applied.

Although the Ising model is successful in explaining some features of our
Monte Carlo simulations, it is incomplete because it assumes perfect nematic
order---the molecules can have only two possible orientations, up and down.
It cannot describe the behavior of the nematic order parameter as a function
of temperature.  For that reason, we proceed to a more general mean-field
theory, in which each molecule has full rotational degrees of freedom.

\subsection{General Mean-Field Calculation}

In mean-field theory, the free energy can be written as
\begin{equation}
F=U-T S=\langle H\rangle+k_B T\langle\log\rho\rangle,
\end{equation}
averaged over the single-particle distribution function $\rho$.  Thus, our
goal is to express the single-particle distribution function in terms of some
variational parameters, calculate the energetic and entropic terms in the free
energy, and then minimize the free energy over those variational parameters.

For this mean-field calculation, we write the molecular orientation at each
lattice site in terms of the polar angle $\theta_i$ and azimuthal angle
$\phi_i$ with respect to the local director.  We assume the distribution
function depends only on the polar angle $\theta_i$, and hence write
\begin{equation}
\label{distfunc}
\rho(\theta_i)=\frac{\exp[v_1 P_1(\cos\theta_i)+v_2 P_2(\cos\theta_i)]}%
{\int_0^\pi \exp[v_1 P_1(\cos\theta_i)+v_2 P_2(\cos\theta_i)]d\theta_i},
\end{equation}
where $v_1$ and $v_2$ are variational parameters.  The order parameters are
then $\langle P_1\rangle =\int_0^\pi P_1(\cos\theta)\rho(\theta)d\theta$
and $\langle P_2\rangle =\int_0^\pi P_2(\cos\theta)\rho(\theta)d\theta$, and
the partition function is
$Z=\int_0^\pi \exp[v_1 P_1(\cos\theta)+v_2 P_2(\cos\theta)]d\theta$.  The
entropic contribution to the free energy is therefore
\begin{equation}
-T S=k_B T\langle\log\rho(\theta_i)\rangle=
k_B T[v_1\langle P_1\rangle+v_2\langle P_2\rangle-\log(Z)]
\end{equation}
per lattice site.

\begin{figure*}
(a)\subfigure{\includegraphics[width=3.0in]{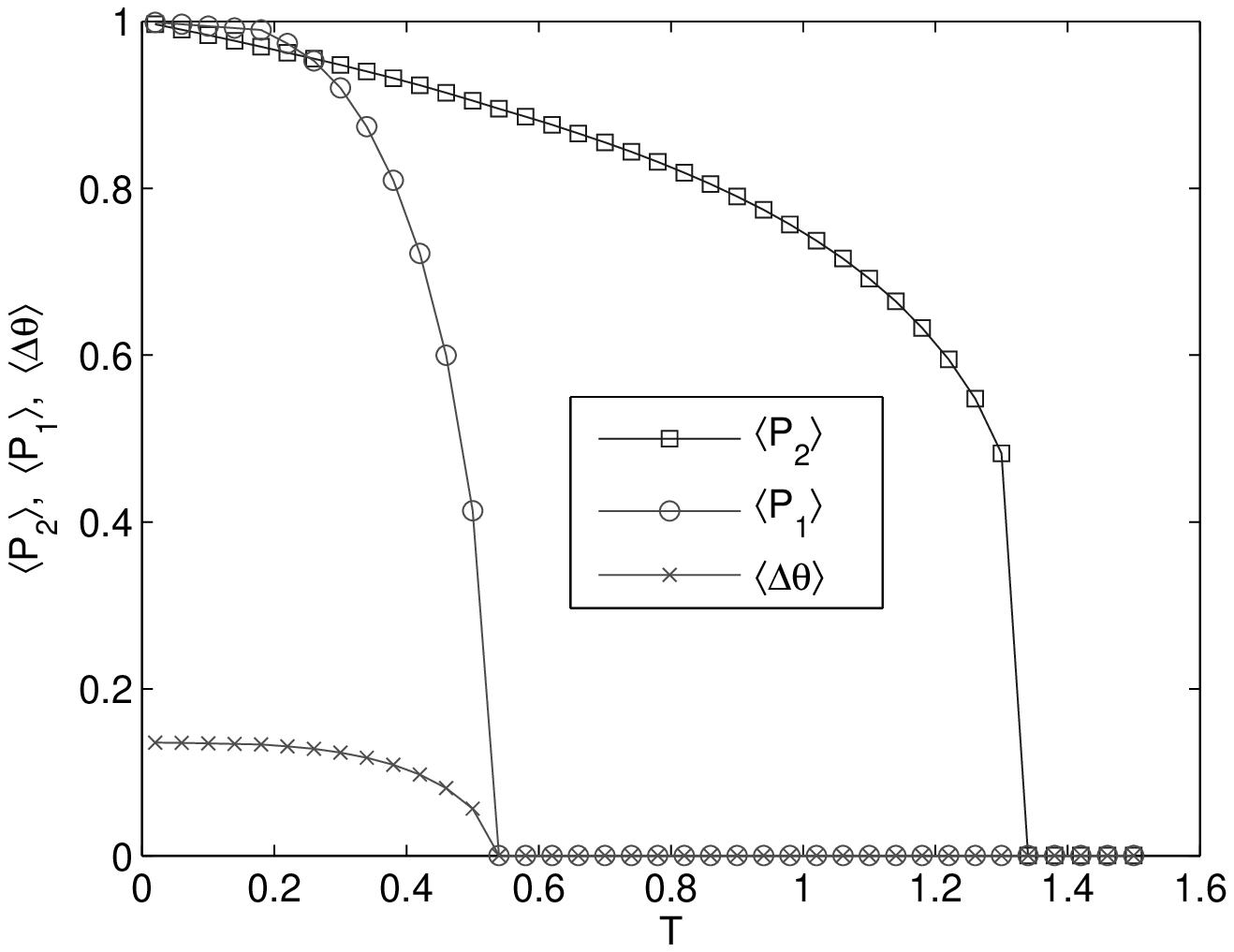}}
(b)\subfigure{\includegraphics[width=3.0in]{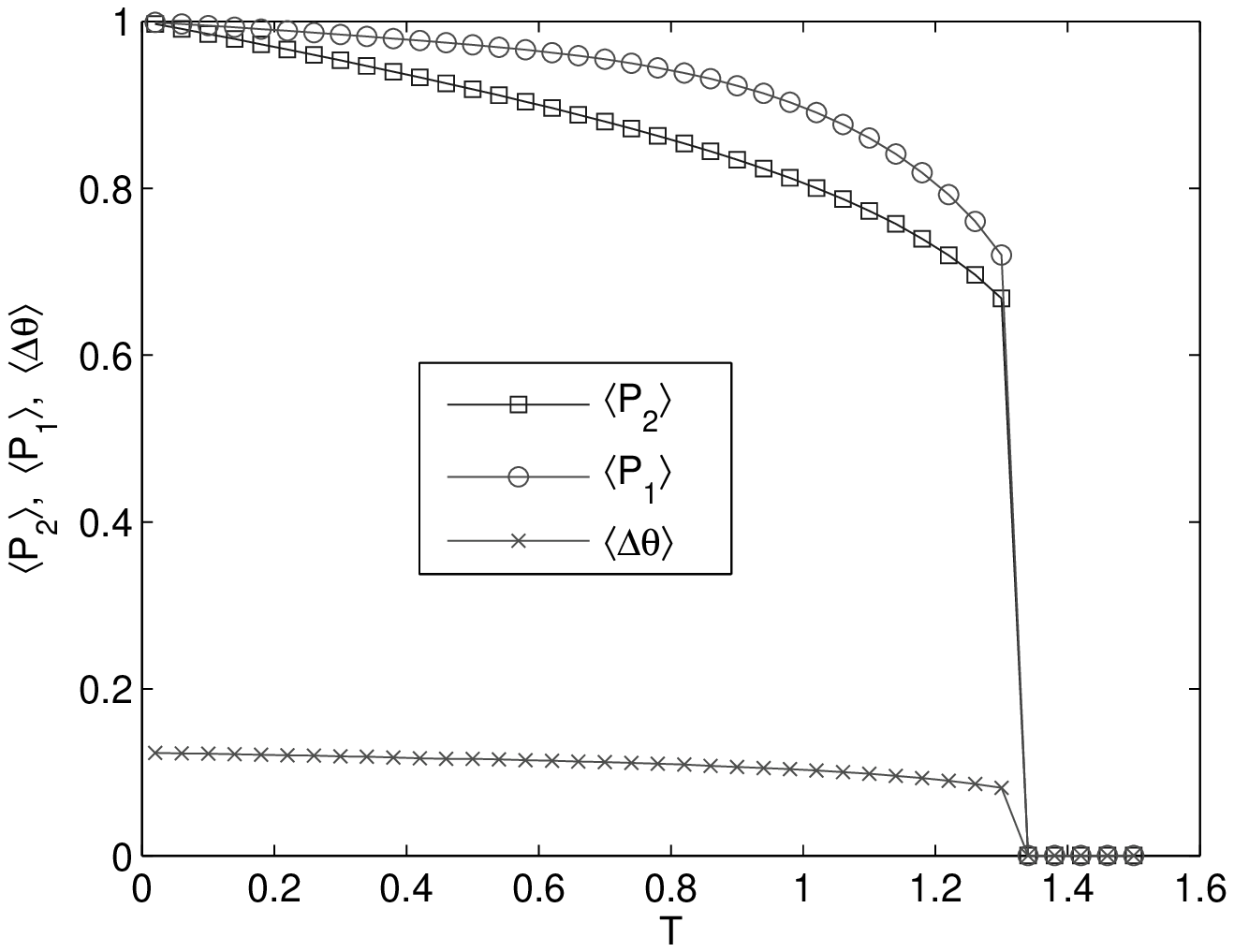}}
(c)\subfigure{\includegraphics[width=3.0in]{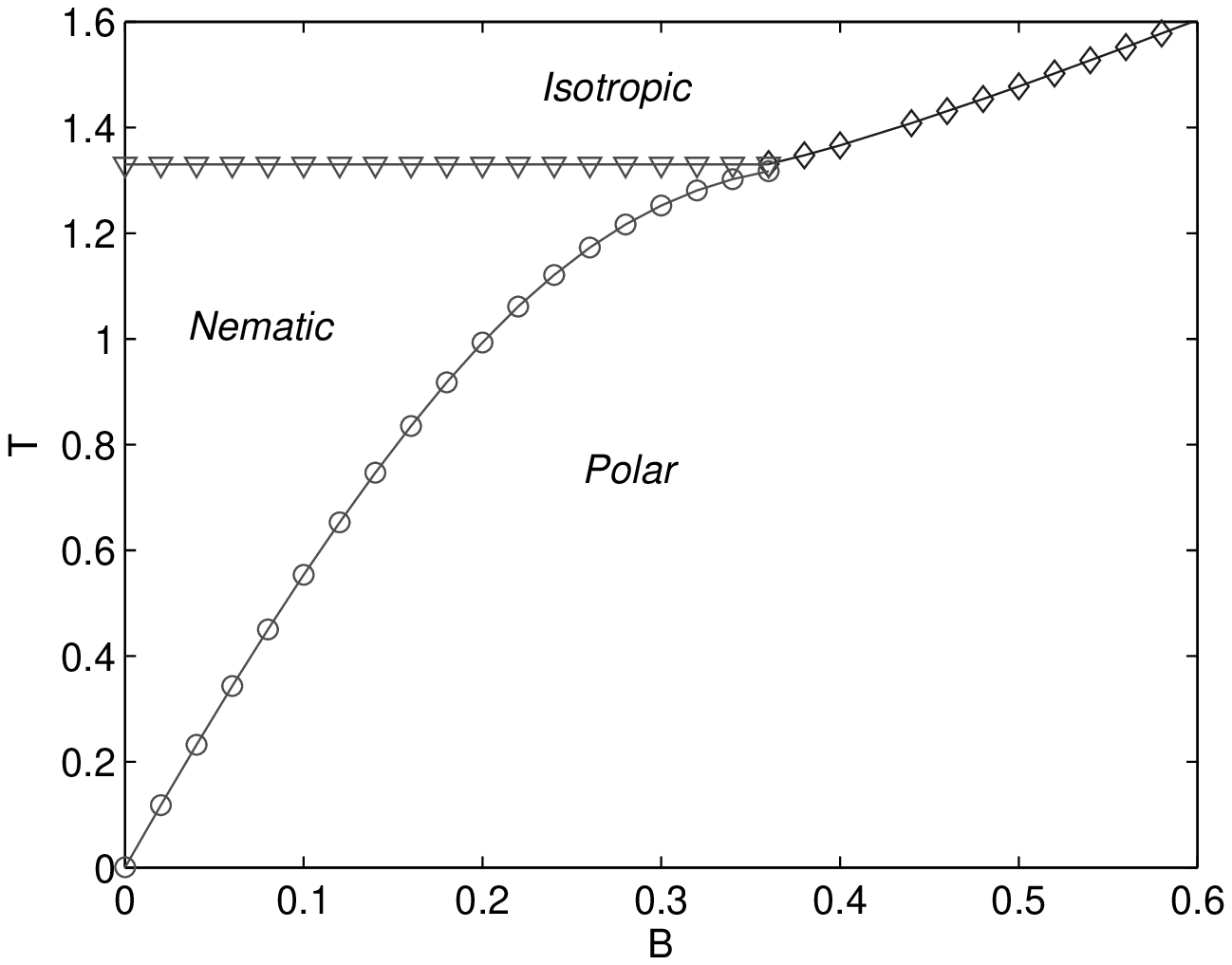}}
(d)\subfigure{\includegraphics[width=3.0in]{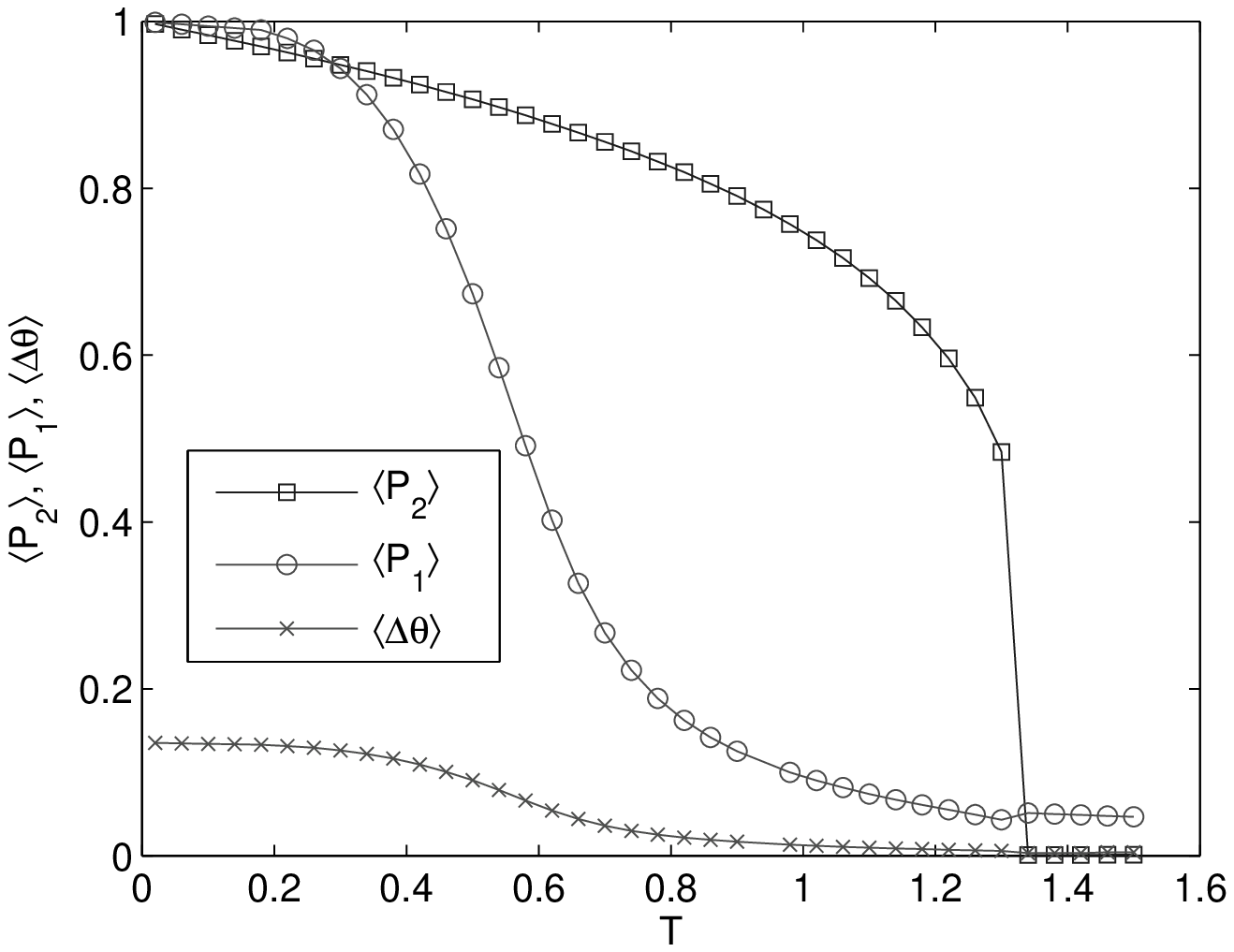}}
\caption{\label{meanfieldplots}Numerical mean-field calculations of the
order parameters $\langle P_1 \rangle$, $\langle P_2 \rangle$, and
$\Delta\theta$ as functions of temperature $T$, for different
values of the interaction parameters chosen to show various types of
transitions:  (a)~Zero-field results for $A=1.5$, $B=0.09$, $C=0.3$, showing
the isotropic-nematic and nematic-polar transitions.  (b)~Zero-field results
for $A=1.5$, $B=0.4$, $C=0.3$, showing the direct isotropic-polar transition.
(c)~Phase diagram for zero field.  (d)~Simulation with applied electric field
$E=0.06$, for the same parameters as in part~(a), showing the induced polar
order and splay in the nematic phase.}
\end{figure*}

As in the previous section, we suppose that site $i$ has its director along
the $z$-axis, as do the two neighbors above and below, while the four nearest
neighbors in the $xy$-plane have directors that are splayed outward by a small
angle $\Delta\theta$.  To calculate the average energy, we combine our
distribution function of Eq.~(\ref{distfunc}) with the lattice Hamiltonian of
Eq.~(\ref{hamiltonian}).  After averaging over all the angles, neglecting
terms involving the third-order Legendre polynomials, and expanding to second
order in the small splay $\Delta\theta$, we obtain
\begin{eqnarray}
\langle H\rangle&=&-A
-A\langle P_2\rangle^2\left[2-2(\Delta\theta)^2\right]\\
&&-B\langle P_1\rangle^2\left[3-(\Delta\theta)^2\right]\nonumber\\
&&-C\langle P_1\rangle\Delta\theta
\left[\frac{5+16\langle P_2\rangle}{15}\right]
-E\langle P_1\rangle\left[1-\frac{(\Delta\theta)^2}{6}\right]\nonumber
\end{eqnarray}
per lattice site.

We now have an expression for the total free energy
\begin{eqnarray}
F&=&-A
-A\langle P_2\rangle^2\left[2-2(\Delta\theta)^2\right]
-B\langle P_1\rangle^2\left[3-(\Delta\theta)^2\right]\nonumber\\
&&-C\langle P_1\rangle\Delta\theta
\left[\frac{5+16\langle P_2\rangle}{15}\right]
-E\langle P_1\rangle\left[1-\frac{(\Delta\theta)^2}{6}\right]\nonumber\\
&&+k_B T[v_1\langle P_1\rangle+v_2\langle P_2\rangle-\log(Z)]
\end{eqnarray}
per lattice site.  In this expression, note that $\langle P_1\rangle$,
$\langle P_2\rangle$, and $Z$ are all determined by the parameters $v_1$ and
$v_2$ in the distribution function.  Hence, the free energy is a function of
just three variational parameters:  $v_1$, $v_2$, and $\Delta\theta$.  Thus,
in the mean-field calculation, we must minimize the free energy numerically
with respect to those three parameters.  After this minimization, we can
calculate the order parameters $\langle P_1\rangle$ and $\langle P_2\rangle$,
and hence determine whether the system is in an isotropic, nematic, or polar
phase.

Figure~\ref{meanfieldplots} shows the numerical mean-field results
for the order parameters $\langle P_1\rangle$ and $\langle P_2\rangle$ and
splay $\Delta\theta$, as well as a complete phase diagram as a
function of temperature $T$.  For a small polar interaction $B=0.09$ there
are two transitions, first from the high-temperature isotropic phase
($\langle P_1\rangle=0$, $\langle P_2\rangle=0$) to the intermediate nematic
phase ($(\langle P_1\rangle=0$, $\langle P_2\rangle\neq 0$), and then from the
nematic phase to the low-temperature polar phase ($\langle P_1\rangle\neq 0$,
$\langle P_2\rangle\neq 0$).  The isotropic-nematic transition is
first-order, while the nematic-polar transition is second-order.  On
increasing the polar interaction strength $B$, the polar phase becomes
stable even at higher temperature. For $B=0.36$, these two transitions
merge into a single first-order transition directly from the high-temperature
isotropic phase to the low-temperature polar phase.  If $E\neq 0$, the
polarization and splay are nonzero even in the nematic phase and scale
with the magnitude of the field.  For nonzero field, the magnitude of the
splay increases on reducing temperature and is enhanced greatly near the
transition to the polar phase.

Note that the numerical mean-field results of Fig.~\ref{meanfieldplots} are
very similar to the Monte Carlo simulation results of
Fig.~\ref{orderparameterplots}, both in the overall phase diagram and in the
splay and polar response to an electric field.  This similarity demonstrates
that the mean-field theory captures the essential physics of this model.

\section{Discussion}

In conclusion, we have developed a lattice model for splay flexoelectricity in
a system of uniaxial pear-shaped molecules.  This model predicts a phase
diagram showing isotropic, nematic, and polar phases, and it further predicts
a converse flexoelectric effect in the nematic phase.  The converse
flexoelectric effect is proportional to the applied electric field, and it
increases dramatically as the temperature decreases toward the nematic-polar
transition.  Indeed, we can regard this effect as a susceptibility to an
applied field, which diverges at the second-order transition to a polar phase.
Thus, flexoelectricity is not just a molecular effect arising from the
microscopic interaction of liquid crystals with a field.  Rather, it is a
statistical effect associated with the response of correlated volumes of
molecules, which increases as one approaches a polar phase.

The recent experiments of Harden et al.~\cite{Harden06,Harden08} have found an
anomalously large \emph{bend} (rather than splay) flexoelectric effect in
systems of bent-core liquid crystals.  We speculate that the same
considerations discussed in this paper can explain the large bend
flexoelectric coefficient in those experiments.  The bent-core liquid crystal
should be close to a polar phase, with order in the transverse dipole moments
of the molecules.  As a result, there should be large correlated volumes of
molecules, leading to a high susceptibility to an applied field, which induces
both polar order and bend.  The detailed theoretical model for bend
flexoelectricity will necessarily be more complex than the model for splay
flexoelectricity presented here, because the bent-core molecules are not
uniaxial and hence their orientations must be characterized by two vectors or
three Euler angles.  This model for bend flexoelectricity will be the subject
of a future paper.

\begin{acknowledgments}
We would like to thank A.~Jakli and R.~L.~B. Selinger for many helpful
discussions. This work was supported by the National Science Foundation
through Grant DMR-0605889.  Computational resources were provided by the Ohio
Supercomputer Center and the Wright Center of Innovation for Advanced Data
Management and Analysis.
\end{acknowledgments}


\begin{thebibliography}{}

\bibitem{meyer69} R. B. Meyer, Phys. Rev. Lett.  \textbf{22}, 918 (1969).

\bibitem{Harden06} J. Harden, B. Mbanga, N. Eber, K. Fodor-Csorba, S. Sprunt,
J. T. Gleeson, and A. Jakli, Phys. Rev. Lett. \textbf{97}, 157802 (2006).

\bibitem{Harden08} J. Harden, R. Teeling, J. T. Gleeson, S. Sprunt, and A.
Jakli, Phys. Rev. E. \textbf{78}, 031702 (2008).

\bibitem{Straley76} J. P. Straley  Phys. Rev. A  \textbf{14}, 1835 (1976).

\bibitem{Osipov84} M. A. Osipov, J. Phys. (Paris) Lett. \textbf{45}, 823
(1984).

\bibitem{Singh89} Y. Singh and U. P. Singh, Phys. Rev. A \textbf{39}, 4254
(1989).

\bibitem{Somoza91} A. M. Somoza and P. Tarazona, Molec. Phys. \textbf{72},
911 (1991).

\bibitem{Zannoni91} F. Biscarini, C. Zannoni, C. Chiccoli, and P. Pasini, Mol.
Phys. \textbf{73}, 439 (1991).

\bibitem{Murthy93} P. R. M. Murthy, V. A. Raghunathan, and N. V. Madhusudana,
Liq. Cryst. \textbf{14}, 483 (1993).

\bibitem{Blinov98} L. M. Blinov, Liq. Cryst. \textbf{24}, 143 (1998).

\bibitem{Stelzer99} J. Stelzer, R. Berardi, and C. Zannoni, Chem. Phys. Lett.
\textbf{299}, 9 (1999).

\bibitem{Billeter00} J. L. Billeter and R. A. Pelcovits, Liq. Cryst.
\textbf{27}, 1151 (2000).

\bibitem{Blinov00} L. M. Blinov, M. I. Barnik, M. Ozaki, N. M. Shtykov, and
K. Yoshino, Phys. Rev. E \textbf{62}, 8091 (2000).

\bibitem{Berardi01} R. Berardi, M. Ricci, and C. Zannoni, ChemPhysChem
\textbf{2}, 443 (2001).

\bibitem{Zannoni01} C. Zannoni, J. Mater. Chem. \textbf{11}, 2637 (2001).

\bibitem{Ferrarini01} A. Ferrarini, Phys. Rev. E \textbf{64}, 021710 (2001).

\bibitem{Dewar05} A. Dewar and P. J. Camp, J. Chem. Phys. \textbf{123}, 174907
(2005).

\bibitem{Kapanowski07} A. Kapanowski, Phys. Rev. E \textbf{75}, 031709 (2007).

\bibitem{Lebwohl72} P. Lebwohl and G. Lasher, Phys. Rev. A \textbf{6}, 426
(1972).

\bibitem{Vsim} http://inac.cea.fr/L\_Sim/V\_Sim/index.en.html.

\bibitem{Geng09} J. Geng and J. V. Selinger, Phys. Rev. E \textbf{80}, 011707
(2009).

\bibitem{Kamien01} For a survey of modulated structures in liquid crystals,
see R. D. Kamien and J. V. Selinger, J. Phys.: Condens. Matter \textbf{13}, R1
(2001).

\end{thebibliography}
\end{document}